 \documentclass[final,5p,times,twocolumn]{elsarticle}

\biboptions{sort&compress}
\usepackage{amssymb}
\usepackage{lipsum}
\usepackage{amsmath}
\usepackage{tikz-feynman}
\tikzfeynmanset{compat=1.1.0}
\usepackage{subcaption}
\usepackage{multirow}
\usepackage{cuted}
\usepackage{hyperref}
\usepackage{ulem}

\journal{Physics Letters B}

\begin{document}

\makeatletter
\def\ps@pprintTitle{%
 \let\@oddhead\@empty
 \let\@evenhead\@empty
 \let\@oddfoot\@empty
 \let\@evenfoot\@empty
}
\makeatother
\begin{frontmatter}



\title{Revisiting the Electron EDM in the NMSSM}


\author[first]{Kaifei Ning}
\ead{kning@umass.edu}
\author[second,third,first,fourth]{Michael Ramsey-Musolf}
\ead{mjrm@sjtu.edu.cn, mjrm@physics.umass.edu}
\affiliation[first]{organization={Amherst Center for Fundamental Interactions, Department of Physics},
            addressline={University of the Massachusetts}, 
            city={Amherst},
            postcode={01003}, 
            state={MA},
            country={USA}}
\affiliation[second]{organization={Tsung-Dao Lee Institute and School of Physics and Astronomy},
            addressline={Shanghai Jiao Tong University}, 
            city={Shanghai},
            postcode={200120}, 
            country={China}} 
\affiliation[third]{Shanghai Key Laboratory for Particle Physics and Cosmology, 
Key Laboratory for Particle Astrophysics and Cosmology (MOE), 
Shanghai Jiao Tong University, Shanghai 200240, China}
\affiliation[fourth]{Kellogg Radiation Laboratory, California Institute of Technology,
Pasadena,
CA 91125, USA}

\begin{abstract}
The Next-to-Minimal Supersymmetric Standard Model (NMSSM) with explicit CP violation offers a promising framework for explaining the observed baryon asymmetry while remaining consistent with stringent electric dipole moment (EDM) bounds. In this work, we identify a previously overlooked set of two-loop diagrams that contribute to fermion EDMs in the NMSSM. Our analysis of the electron EDM shows that, for certain NMSSM specific CP-violating phases, these diagrams partially cancel existing contributions, thereby relaxing the associated constraints. In other phases, the diagrams dominate and tighten the bounds, which in turn necessitates more finely tuned parameters to reconcile successful baryogenesis with current EDM limits.
\end{abstract}



\end{frontmatter}



\section{Introduction}
\label{sec:introduction}
Explaining the origin of the observed baryon asymmetry of the universe (BAU)\cite{ParticleDataGroup:2024cfk} remains a central challenge in cosmology and particle physics. Although the Standard Model (SM) contains the essential ingredients for baryogenesis, it fails to produce a sufficiently strong first-order phase transition and enough CP violation to account for the observed matter–antimatter imbalance~\cite{Cohen:1993nk,Kajantie:1995kf}. 
Historically, extensions beyond the SM, such as the Minimal Supersymmetric Standard Model (MSSM), were considered promising frameworks to satisfy the Sakharov conditions. The MSSM introduces new potential sources of CP violation and mechanisms involving scalar fields (like light top squarks) that could, in principle, induce a strong first-order EWPT needed for electroweak baryogenesis (EWBG) \cite{Delepine:1996vn,Huet:1995sh,Cline:1997vk,Carena:1996wj,Cline:2000nw,Lee:2004we,Carena:2008rt}.

Despite these appealing features, realizing successful EWBG within the MSSM faces significant hurdles, casting doubt on its viability as a simple solution. Theoretically, the model suffers from the $\mu$-problem that the supersymmetric parameter $\mu$ must be finely tuned before spontaneous symmetry breaking to generate the correct electroweak scale~\cite{Kim:1983dt}, raising naturalness concerns. Experimentally, achieving a strong first-order EWPT typically requires light stop quarks $m_{\tilde{t}}\lesssim m_t$, but searches at the Large Hadron Collider (LHC) have pushed the lower limits on stop masses significantly, now exceeding 1 TeV~\cite{CMS:2021xyz}, creating strong tension with this requirement. Furthermore, recent electric dipole moment (EDM) measurements impose stringent constraints on CP-violating interactions, significantly reducing the viable MSSM parameter space~\cite{Chupp:2017rkp,ACME:2018yjb,Panico:2018hal,Roussy:2022cmp,Bodeker:2020ghk}. 

Possible mechanisms have been proposed to address these difficulties. For example, Ref.~\cite{Ross:2016pml} has introduced a non-holomorphic soft term $\mu' \tilde{H}_d \tilde{H}_u$, which allows for a heavier Higgsino and alleviates the fine-tuning of $\mu$. Refs.~\cite{Li:2010ax,Han:2021ify,Nakai:2016atk,Cesarotti:2018huy} suggest large CP-violating phases could survive EDM constraints, although such scenarios require highly fine-tuned configurations. The most compelling solution, however, lies in the Next-to-Minimal Supersymmetric Standard Model (NMSSM), which incorporates an additional singlet superfield coupled to the Higgs fields via $\lambda \hat{S} \hat{H}_d \hat{H}_u$~\cite{Fayet:1974fj,Fayet:1974pd,Fayet:1977yc,Ellis:1988er,Ellwanger:1993xa}. This interaction resolves the $\mu$-problem, as it generates an effective $\mu_{\text{eff}} = \lambda \langle S \rangle$ at the desired weak scale when the scalar component of the singlet $S$ acquires a vacuum expectation value (VEV). Additionally, the extra degree of freedom in the scalar spectrum helps induce a strong first-order phase transition without relying on light stops. Notably, the NMSSM introduces more CP-violating interactions, both in the superpotential and soft-breaking terms, thus opening new avenues for baryogenesis and EDM analysis. Comprehensive reviews of the NMSSM can be found in Refs.~\cite{Ellwanger:2009dp,Maniatis:2009re}. 

Previous studies have largely focused on the $\mathbb{Z}_3$-invariant NMSSM, that is, NMSSM with a discrete $\mathbb{Z}_3$ symmetry. In this scenario, there 
exists one additional independent CP-violating phase beyond those in the MSSM,
$\phi_3$, along with two extra real scalars $(h_s, a_s)$ and an additional neutralino $\tilde{S}$ beyond the MSSM field content. Building on prior MSSM-based 
analyses, Ref.~\cite{Cheung:2011wn} was the first to present analytic expressions for the dominant fermion EDM contributions, including the one-loop neutralino exchange diagram $(d_f^E)^{\tilde{\chi}^0}$~\cite{Ibrahim:1997gj} and the two-loop 
Barr--Zee diagrams $(d_f^E)^{\gamma H}$, $(d_f^E)^{W H}$, $(d_f^E)^{W W}$, and $(d_f^E)^{Z H}$, generated by effective $\gamma\gamma H_i^0$, $\gamma H^\pm W^\mp$, $\gamma W^\pm W^\mp$, and $\gamma H_i^0 Z$ operators~\cite{Barr:1990vd,Chang:2005ac,Li:2008kz,Ellis:2008zy}, respectively. The one-loop chargino contribution $(d_f^E)^{\tilde{\chi}^\pm}$ 
was neglected since $\phi_3$ has a negligible impact on it. Including the effects of chromo-electric dipole moment (CEDM) of light quarks, renormalization group mixing with the three-gluon operator, and threshold corrections from CP-violating four-quark interactions, the aforementioned studies found that even a large CP phase of $\phi_3 \sim 90^\circ$ could remain consistent with then-current experimental bounds on the thallium, neutron, mercury, deuteron, and radium EDMs. Following a similar methodology, Ref.~\cite{King:2015oxa} examined the parameter space for both NMSSM- and MSSM-specific CP phases allowed by LHC Higgs measurements and EDM constraints. The results indicate that the NMSSM-specific CP-violating phase can be more strongly limited by Higgs data than by EDM searches. Recently, Ref.~\cite{Dao:2022rui} emphasizes 
that large cancellations in the electron EDM are required to satisfy the current experimental upper bound.

In this work, instead of restricting to $\mathbb{Z}_3$-invariant interactions we consider the NMSSM with the most general renormalizable couplings as in Ref.~\cite{Ellwanger:2009dp}. Certain terms, such as $\beta \hat{S}^2$, conflict with the motivation to reduce fine-tuning; hence, a non-holomorphic soft term $\mu^\prime \tilde{H}_d \tilde{H}_u$ is introduced in the soft breaking Lagrangian to address this issue. We present a complete basis of CP-violating phases invariant under various global $U(1)$ transformations, provided appropriate spurion charges to NMSSM fields are introduced. Within the effective field theory framework, these phases generate various dimension-6 CP-violating operators, such as fermion EDMs, quark chromo-EDMs, the three-gluon Weinberg operator, and four-fermion interactions. The dimension-4 QCD $\bar{\theta}$ term can also receive contributions, not only through radiatively induced quark masses but also via threshold corrections from dimension-6 operators such as the quark CEDM~\cite{Pospelov:2005pr,Engel:2013lsa}. Such effects could reintroduce a dangerously large $\bar{\theta}$ value unless dynamically suppressed, for instance, by the Peccei-Quinn (PQ) mechanism. Consequently, these operators induce low-energy observables, including EDMs of electrons, nucleons, atoms, and molecules. Our work specifically targets the electron EDM, which currently provides the most sensitive probe of CP violation in BSM physics and imposes the strongest constraints on the relevant phases. The electron EDM receives contributions from these phases through their appearance in unitary rotation matrices and CP-violating couplings in the mass eigenbasis. Of particular relevance, we identify three classes of two-loop diagrams $(d_f^E)^{H^\pm}$, $(d_f^E)^{W^\pm}$, and $(d_f^E)^{\text{Kite}}$ that contribute to fermion EDMs in the NMSSM but have been overlooked in the literature. These diagrams share topological similarities with those in the two-Higgs-doublet model (2HDM) and are absent in the MSSM due to the lack of tree-level CP violation in its scalar sector. However, they must be included in the NMSSM or any MSSM extension with an enlarged Higgs sector, such as 
the $\mu\nu$SSM~\cite{Lopez-Fogliani:2005vcg, Lopez-Fogliani:2020gzo}. We generalize the analytic expressions for fermion EDMs from the 2HDM to the NMSSM framework. Under a specific benchmark choice, we compare the magnitudes of these diagrammatic contributions and numerically study the constraints on each CP-violating phase imposed by the current electron EDM bound, $d_e \le 4.1 \times 10^{-30}\, \mathrm{e\,cm}$~\cite{Roussy:2022cmp}. We further investigate the impact of these additional two-loop diagrams on the NMSSM-specific CP phases. 

In Table~\ref{Tab:DiagramPhaseDependence}, we summarize the dependence of the leading EDM diagrams on various CP-violating phases, grouping them into three categories: one-loop MSSM (MSSM~OL), two-loop MSSM (MSSM~TL), and 2HDM-type diagrams. Each type of diagram yields a gauge-independent contribution to the electron EDM. Comparing the sensitivities of each category to the invariant phases in Fig.~\ref{fig:singlephasecompare}, we observe that  $\phi_{3,4,5}$ contribute primarily via the 2HDM-type diagrams, whereas for $\phi_0^\prime$ and $\phi_6$, 2HDM corrections are subdominant and opposite in sign compared to the MSSM~TL. Other phases, such as 
$\Phi_1, \phi_{0,2,3,9}$, remain largely unaffected by 2HDM diagrams and instead contribute at roughly the same order of magnitude to MSSM~OL and MSSM~TL. From Fig.~\ref{fig:singlephaseconstraint}, we find that $\Phi_{3,4}, \phi_0^\prime, \phi_{1,6}$ are subject to weaker constraints from the electron EDM than $\Phi_1, \phi_{0,2,3,4,5,9}$. The former set can evade the electron EDM bound with phases as large as $\mathcal{O}(10^{-1})$, while the magnitudes of the latter are restricted to be below $\mathcal{O}(10^{-2})$ or even $\mathcal{O}(10^{-3})$. We additionally perform a numerical scan over $\phi_{3,4,5}$ and show that including or excluding 2HDM diagrams significantly alters the allowed parameter space of these phases (see Fig.~\ref{fig:scan}). In particular, $\phi_4$ is more tightly constrained by 2HDM terms than $\phi_{3}$ or $\phi_5$, and a fine-tuned cancellation between $\phi_4$ and $\phi_5$ is necessary to remain compatible with the current electron EDM limit.

Our letter is organized as follows. In Section~2, we present the relevant Lagrangian terms and introduce the complete set of invariant CP-violating phases that appear in the mass spectrum and electron EDM. We then outline the EDM diagrams included in our analysis and discuss the numerical results in Section~3. Section~4 provides a summary of our findings and conclusions.

\section{Theoretical framework}
\label{sec:model building}
\subsection{The general NMSSM}

The general NMSSM is defined as MSSM extended with a gauge singlet chiral superfield $\hat{S}$. The most general superpotential with all possible renormalizable interactions takes the form,
\begin{equation}
    \label{eq:Superpotential}
    \begin{aligned}
        W^{}
        &= \lambda\hat{S}\hat{H}_u\hat{H}_d + \frac{\kappa}{3}\hat{S}^3 + \frac{1}{2}\beta\hat{S}^2 + \alpha\hat{S} + W^{\text{MSSM}}
    \end{aligned}
\end{equation}
The corresponding soft SUSY-breaking Lagrangian is
\begin{equation}
    \label{eq:softLagrangian}
    \begin{aligned}
        \mathcal{L}_{\text{soft}}^{} 
        =& - \bigl[\lambda A_\lambda S H_u H_d + \frac{\kappa A_\kappa}{3}S^3 + \frac{1}{2}m_7^2S^2 + m_9^3S + h.c.\bigr]\\
        &- m_S^2 S^\dagger S- \mu^\prime \tilde{H}_d \tilde{H}_u  + \mathcal{L}_{\text{soft}}^{\text{MSSM}}\\
    \end{aligned}
\end{equation}
As we do not impose a discrete $\mathbb{Z}_3$ symmetry on the theory, an explicit $\mu \hat{H}_u \hat{H}_d$ term remains in the MSSM sector of the superpotential, thereby conflicting with the NMSSM's original aim of resolving the ``$\mu$-problem''. To mitigate this issue, we introduce a non-holomorphic soft term, $\mu^\prime \tilde{H}_d \tilde{H}_u$, which helps alleviate the fine-tuning associated with $\mu$~\cite{Ross:2016pml}. We  expand the Higgs doublets and the singlet field around their vacuum expectation values (VEVs) as follows:
\begin{equation}
\label{eq:HiggsExpansion}
\begin{aligned}
    H_u^T &= \biggl(H_u^+,\frac{v_u + h_u + ia_u}{\sqrt{2}}\biggr)e^{i\varphi_u},\\
    H_d^T &= \biggl(\frac{v_d + h_d + ia_d}{\sqrt{2}},H_d^-\biggr)e^{i\varphi_d},\\
    S &= \frac{v_s + h_s + ia_s}{\sqrt{2}}e^{i\varphi_s}
\end{aligned}
\end{equation}
Taking first derivatives of the Higgs potential and imposing the conditions 
\(\partial V/\partial h_{d,u,s} = 0\) and \(\partial V/\partial a_{d,u,s} = 0\) 
at the VEVs yields six tadpole equations. At tree level, the first three equations can be used to solve for the soft masses \(m^2_{H_u}\), \(m^2_{H_d}\), and \(m^2_S\). Meanwhile, the equations 
\(\partial V/\partial a_{d,u} = 0\) are degenerate, so only two imaginary components of the Higgs-sector couplings can be determined when combined with \(\partial V/\partial a_{s} = 0\).

Exploiting global symmetries \(U(3)^5\), \(U(1)_{PQ}\), and \(U(1)_R\) allows one to remove multiple nonphysical CP-violating phases (see Refs.~\cite{Chung:2003fi,Haber:1997if}) In the NMSSM, introducing the singlet 
superfield \(\hat{S}\) also permits constructing a new \(U(1)_S\) symmetry, under which the scalar \(S\) and the spinor \(\tilde{S}\) each carry charge \(+1\). Consequently, every singlet-related coupling acquires a corresponding spurion \(U(1)_S\) charge. Assuming flavor-diagonal soft mass matrices \(\mathbf{M}^2_{\tilde{Q},\tilde{U},\tilde{D},\ldots}\) and trilinear couplings \(A_{u,d,e}\) aligned with the Yukawa matrices, one obtains a basis of 16 CP-violating phases after performing appropriate field rephasings and shifts. Half of these phases originate from the MSSM sector, 
\begin{equation}
\label{eq:MSSMphases}
   \begin{aligned}
       &\Phi_1 = \arg\{M_1M_2^\ast\}, \quad \Phi_2 = \arg\{M_1M_3^\ast\}, \quad \Phi_3 = \arg\{A_uA_e^\ast\},\\ 
    &\Phi_4 = \arg\{A_dA_e^\ast\},\quad \Phi_5 = \theta_{CKM}, \quad \phi_0 = \arg\{bz_u z_d\}, \\ &\phi_1 = \arg\{M_1A_e^\ast\}, \quad \phi_9 = \arg\{M_1 \mu^\prime b^\ast\},
   \end{aligned}
\end{equation}
Here and throughout, $z_u, z_d,$ and $z_s$ denote \(\langle H_u^0\rangle, \langle H_d^0\rangle,\) and \(\langle S\rangle,\) respectively. The phases \(\Phi_{1,2,3,4}\) and \(\phi_1\) originate from 
complex gaugino and sfermion couplings, whereas \(\Phi_5\) is tied to the SM sector. The phase \(\phi_0\) vanishes in the pure MSSM due to its minimization conditions but acquires a non-trivial value in the NMSSM. Meanwhile, \(\phi_9\) arises from the non-holomorphic term. The remaining half of the phases are associated with NMSSM-specific couplings,
\begin{equation}
\label{eq:NMSSMphases}
\begin{aligned}
    \phi_0^\prime &= \arg(\kappa A_\kappa z_s^3), &\phi_2 = \arg(M_1 \lambda z_s b^\ast)\\
    \phi_3 &= \arg(\lambda  z_u z_d \kappa^\ast z_s^{\ast 2}), &\phi_4 = \arg(\lambda A_\lambda z_u z_d z_s)\\
    \phi_5 &= \arg(\lambda z_u z_d \beta^\ast z_s^\ast),&\phi_6 = \arg(m_7^2 z_s^2)\\
    \phi_7 &= \arg(m_9^3 z_s), &\phi_8 = \arg(\lambda z_u z_d \alpha^\ast)
\end{aligned}
\end{equation} 
where the phases $\phi_0^\prime$, $\phi_3$, and $\phi_4$ appear in $\mathbb{Z}_3$-NMSSM, while $\phi_{2,5,6,7,8}$ arise from the quadratic and linear interactions detailed in Eqs.~\eqref{eq:Superpotential} and \eqref{eq:softLagrangian}. In addition, $\phi_2$ corresponds to an effective form of the MSSM invariant $\arg(M_1 \mu b^\ast)$, with $\lambda z_s = \mu_{\mathrm{eff}}$. In our chosen basis, $\arg(M_1 \mu b^\ast)$ vanishes because the parameter $\mu$ has been absorbed into $\hat{S}$ via a field redefinition. Crucially, not all eight CP-violating phases in the NMSSM sector are independent: two are constrained by the scalar potential minimization conditions, reducing the total number of physical phases to (8+6)=14.
\begin{table}[b]
\begin{tabular}{c c c c} 
 \hline
 Source & Diagram & MSSM Phase & NMSSM Phase \\ 
 \hline
 \multirow{2}{4.5em}{MSSM OL} & $(d_f^E)^{\tilde{\chi}^0}$ & $\Phi_1,~ \phi_{0,1,2,9}$  & $\phi_{3,5}$ \\ 
  & $(d_f^E)^{\tilde{\chi}^\pm}$ & $\Phi_1,~\phi_{0,2,9}$ & 	 \\ 
 \hline
 \multirow{4}{4.5em}{MSSM TL} & $(d_f^E)^{\gamma H}$& $\Phi_{1,3,4}, ~\phi_{0,1,2,9}$ & $\phi_0^\prime, ~\phi_{3,4,5,6,7,8}$ \\ 
  & $(d_f^E)^{Z H}$ & $\Phi_{1}, ~\phi_{0,2,9}$ & $\phi_0^\prime, ~\phi_{3,4,5,6,7,8}$ \\
  & $(d_f^E)^{W H}$ & $\Phi_1,~ \phi_{0,2,9}$  & $\phi_{3,5}$ \\ 
  & $(d_f^E)^{W W}$ & $\Phi_1,~ \phi_{0,2,9}$  & $\phi_{3,5}$ \\
 \hline 
 \multirow{3}{4.5em}{2HDM}  & $(d_f^E)^{H^\pm}$ & $\phi_0$ & $\phi_0^\prime, ~\phi_{3,4,5,6,7,8}$ \\ 
  & $(d_f^E)^{W^\pm}$ & $\phi_0$  & 	$\phi_0^\prime, ~\phi_{3,4,5,6,7,8}$ \\ 
  & $(d_f^E)^{\text{Kite}}$ & $\phi_0$  & $\phi_0^\prime, ~\phi_{3,4,5,6,7,8}$ \\
 \hline 
\end{tabular}
\caption{Diagrams that contribute to electron EDM in the NMSSM and their phase dependence. $\Phi_1, \phi_{0,2,9}$ are phases of chargino mass matrix, $\phi_{3,5}$ enter the neutralino mass matrix, $\phi_0^\prime, \phi_{0,3,4,5,6,7,8}$ determine CP-violations of the scalar Higgs sector, and $\Phi_{3,4}, \phi_1$ live in sfermion mass matrices.}
\label{Tab:DiagramPhaseDependence}
\end{table}

\subsection{Electron EDM}
In the pursuit of discovering BSM sources of CP violation, as well as CP violation arising from the QCD $\theta$-term, EDM searches have emerged as some of the most sensitive probes (see, {\it e.g.}, Ref.~\cite{Chupp:2017rkp}). In the (N)MSSM, the electron EDM receives one-loop contributions, which are typically suppressed by heavy sfermion masses. In contrast, certain two-loop diagrams can dominate despite the extra loop factor, particularly the Barr--Zee--type diagrams that involve lighter Higgs or gauge bosons. These two-loop contributions are often classified by the structure of the three-point vertex entering the Barr--Zee diagram. The commonly studied one- and two-loop diagrams for the MSSM and NMSSM, which share analogous topologies, are illustrated in Fig.~\ref{fig:MSSMOneLoop} and Fig.~\ref{fig:MSSMTwoLoop}.

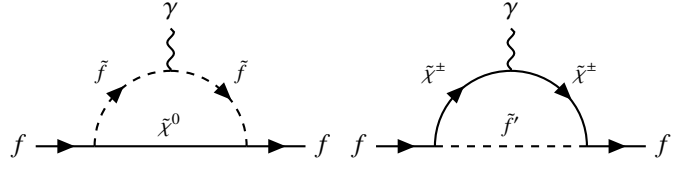
\begin{figure}[t]
    \begin{subfigure}{0.18\textwidth}
        \centering
        \begin{tikzpicture}
        \begin{feynman}
               \vertex at (0,0) (a) {\(f\)};
               \vertex at (1,0) (x);
               \vertex at (2,1) (z);
               \vertex at (2, 1.8) (c) {\(\gamma\)};
               \vertex at (3, 0) (y) ;
               \vertex at (4, 0) (b) {\(f\)};
               \diagram* {
               (a) -- [fermion, thick] (x) -- [edge label = {\footnotesize\(\tilde{\chi}^0\)},thick] (y) -- [fermion,thick] (b),
               (x) -- [charged scalar, quarter left, edge label = {\footnotesize\(\tilde{f}\)}, thick] (z) -- [charged scalar, quarter left, edge label = {\footnotesize\(\tilde{f}\)}, thick] (y),
               (z) -- [photon, thick] (c),
               };
        \end{feynman}
        \end{tikzpicture}
    \end{subfigure}
    \hspace{1cm}
    \begin{subfigure}{0.18\textwidth}
        \centering
        \begin{tikzpicture}
        \begin{feynman}
               \vertex at (0,0) (a) {\(f\)};
               \vertex at (1,0) (x);
               \vertex at (2,1) (z);
               \vertex at (2, 1.8) (c) {\(\gamma\)};
               \vertex at (3, 0) (y) ;
               \vertex at (4, 0) (b) {\(f\)};
               \diagram* {
               (a) -- [fermion, thick] (x) -- [scalar, edge label = {\footnotesize\(\tilde{f}^\prime\)}, thick] (y) -- [fermion, thick] (b),
               (x) -- [fermion, quarter left, edge label = {\footnotesize\(\tilde{\chi}^{\pm}\)}, thick] (z) -- [fermion, quarter left, edge label = {\footnotesize\(\tilde{\chi}^{\pm}\)}, thick] (y),
               (z) -- [photon, thick] (c),
               };
        \end{feynman}
        \end{tikzpicture}
    \end{subfigure}
    \caption{One-loop diagrams contributing to fermion EDM in the MSSM; left: neutralino exchange diagram $(d_f^E)^{\tilde{\chi}^0}$, right: chargino exchange diagram $(d_f^E)^{\tilde{\chi}^\pm}$.}
    \label{fig:MSSMOneLoop}
\end{figure}
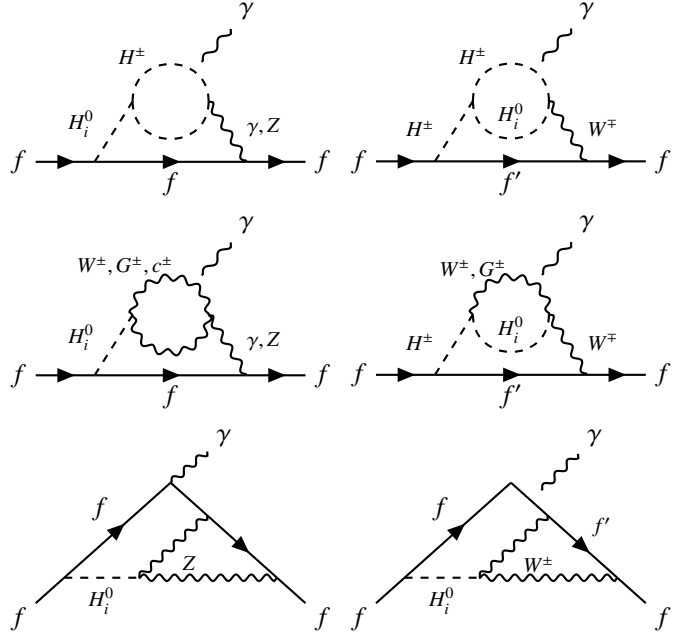
\begin{figure}[t]
    \begin{subfigure}{0.18\textwidth}
        \centering
        \begin{tikzpicture}
        \begin{feynman}
               \vertex at (0,0) (a) {\(f\)};
               \vertex at (1,0) (x);
               \vertex at (3, 0) (y);
               \vertex at (4, 0) (b) {\(f\)};
               \vertex at (1.5, 0.8) (m);
               \vertex at (2.5, 0.8) (n);
               \vertex at (2.3,1.2) (d) {};
               \vertex at (3, 2) (c) {\(\gamma\)};
           \diagram*{
               (a) -- [fermion, thick] (x) -- [fermion, edge label' = \(f\), thick] (y) -- [fermion, thick] (b),
               (x) -- [scalar, thick, edge label ={\footnotesize\(H_i^0\)}, near start] (m),
               (y) -- [boson, thick, edge label'= {\footnotesize\(\gamma,Z\)}, near start] (n),
               (m) -- [scalar, half right, thick, looseness = 1.7] (n) -- [scalar, half right, thick, looseness = 1.7] (m),
               (d) -- [photon, thick] (c),
               };
        \end{feynman}
        \node at (1.5,1.4) {\footnotesize$ H^\pm$};
        \end{tikzpicture}
    \end{subfigure}
    \hspace{1cm}
    \begin{subfigure}{0.18\textwidth}
        \centering
        \begin{tikzpicture}
        \begin{feynman}
               \vertex at (0,0) (a) {\(f\)};
               \vertex at (1,0) (x);
               \vertex at (3, 0) (y);
               \vertex at (4, 0) (b) {\(f\)};
               \vertex at (1.5, 0.8) (m);
               \vertex at (2.5, 0.8) (n);
               \vertex at (2.3,1.2) (d) {};
               \vertex at (3, 2) (c) {\(\gamma\)};
           \diagram*{
               (a) -- [fermion, thick] (x) -- [fermion, edge label' = \(f^\prime\), thick] (y) -- [fermion, thick] (b),
               (x) -- [scalar, thick, edge label ={\footnotesize\(H^\pm\)}, near start] (m),
               (y) -- [boson, thick, edge label'= {\footnotesize\(W^\mp\)}, near start] (n),
               (m) -- [scalar, half right, thick, looseness = 1.7] (n) -- [scalar, half right, thick, looseness = 1.7] (m),
               (d) -- [photon, thick] (c),
               };
        \end{feynman}
        \node at (1.5,1.4) {\footnotesize$ H^\pm$};
        \node at (2,0.6) {\footnotesize$ H_i^0$};
        \end{tikzpicture}
    \end{subfigure} \\
    \begin{subfigure}{0.18\textwidth}
        \centering
        \begin{tikzpicture}
        \begin{feynman}
               \vertex at (0,0) (a) {\(f\)};
               \vertex at (1,0) (x);
               \vertex at (3, 0) (y);
               \vertex at (4, 0) (b) {\(f\)};
               \vertex at (1.5, 0.8) (m);
               \vertex at (2.5, 0.8) (n);
               \vertex at (2.3,1.2) (d) {};
               \vertex at (3, 2) (c) {\(\gamma\)};
           \diagram*{
               (a) -- [fermion, thick] (x) -- [fermion, edge label' = \(f\), thick] (y) -- [fermion, thick] (b),
               (x) -- [scalar, thick, edge label ={\footnotesize\(H_i^0\)}, near start] (m),
               (y) -- [boson, thick, edge label'= {\footnotesize\(\gamma,Z\)}, near start] (n),
               (m) -- [boson, half right, thick, looseness = 1.7] (n) -- [boson, half right, thick, looseness = 1.7] (m),
               (d) -- [photon, thick] (c),
               };
        \end{feynman}
        \node at (1.4,1.45) {\footnotesize$W^\pm, G^\pm, c^\pm$};
        \end{tikzpicture}
    \end{subfigure}
    \hspace{1cm}
    \begin{subfigure}{0.18\textwidth}
        \centering
        \begin{tikzpicture}
        \begin{feynman}
               \vertex at (0,0) (a) {\(f\)};
               \vertex at (1,0) (x);
               \vertex at (3, 0) (y);
               \vertex at (4, 0) (b) {\(f\)};
               \vertex at (1.5, 0.8) (m);
               \vertex at (2.5, 0.8) (n);
               \vertex at (2.3,1.2) (d) {};
               \vertex at (3, 2) (c) {\(\gamma\)};
           \diagram*{
               (a) -- [fermion, thick] (x) -- [fermion, edge label' = \(f^\prime\), thick] (y) -- [fermion, thick] (b),
               (x) -- [scalar, thick, edge label ={\footnotesize\(H^\pm\)}, near start] (m),
               (y) -- [boson, thick, edge label'= {\footnotesize\(W^\mp\)}, near start] (n),
               (m) -- [scalar, half right, thick, looseness = 1.7] (n) -- [boson, half right, thick, looseness = 1.7] (m),
               (d) -- [photon, thick] (c),
               };
        \end{feynman}
        \node at (1.5,1.4) {\footnotesize$W^\pm, G^\pm$};
        \node at (2,0.6) {\footnotesize$ H_i^0$};
        \end{tikzpicture}
    \end{subfigure} \\
\begin{subfigure}{0.18\textwidth}
        \centering
        \begin{tikzpicture}
        \begin{feynman}
               \vertex at (0,0) (a) {\(f\)};
               \vertex at (4, 0) (b) {\(f\)};
               \vertex at (2, 1.8) (c);
               \vertex at ($(a)!0.3!0:(c)$) (m);
               \vertex at ($(b)!0.3!0:(c)$) (n);
               \vertex at ($(m)!0.35!0:(n)$) (x);
               \vertex at ($(b)!0.76!0:(c)$) (y);
               \vertex at (2.7, 2.4) (d) {\(\gamma\)};
           \diagram*{
               (a) -- [thick] (m) -- [fermion, edge label = \(f\), thick] (c) -- [thick] (y) -- [fermion, thick] (n) -- [ thick] (b),
               (m) -- [scalar, thick, edge label' ={\footnotesize\(H_i^0\)}] (x),
               (x) -- [boson, thick, edge label'= {\footnotesize\(Z\)}] (y),
               (x) -- [boson, thick] (n),
               (c) -- [photon, thick] (d),
               };
        \end{feynman}
        \end{tikzpicture}
    \end{subfigure}
    \hspace{1cm}
    \begin{subfigure}{0.18\textwidth}
        \centering
        \begin{tikzpicture}
        \begin{feynman}
               \vertex at (0,0) (a) {\(f\)};
               \vertex at (4,0) (b) {\(f\)};
               \vertex at (2, 1.8) (c);
               \vertex at ($(a)!0.3!0:(c)$) (m);
               \vertex at ($(b)!0.3!0:(c)$) (n);
               \vertex at ($(m)!0.35!0:(n)$) (x);
               \vertex at ($(b)!0.76!0:(c)$) (y);
               \vertex at (2.4, 1.7) (e);
               \vertex at (3.1, 2.4) (d) {\(\gamma\)};
           \diagram*{
               (a) -- [thick] (m) -- [fermion, edge label = \(f\), thick] (c) -- [thick] (y) -- [fermion, thick, edge label ={\footnotesize\(f^\prime\)}] (n) -- [ thick] (b),
               (m) -- [scalar, thick, edge label' ={\footnotesize\(H_i^0\)}] (x),
               (x) -- [boson, thick, edge label'= {\footnotesize\(W^\pm\)}] (y),
               (x) -- [boson, thick] (n),
               (e) -- [photon, thick] (d),
               };
        \end{feynman}
        \end{tikzpicture}
    \end{subfigure}
    \caption{Two-loop diagrams contributing to fermion EDMs in the background field gauge classified as 2HDM contribution; top: charged Higgs loop diagrams $(d_f^E)^{H^\pm}$, middle: W boson loop diagrams $(d_f^E)^{W^\pm}$, bottom: kite diagrams $(d_f^E)^{\text{Kite}}$. $c$ denote ghost fields and the detached photon lines can be inserted in internal charged propagators or allowed vertices. For kite diagrams, other arrangements of $H_i^0$ and $Z/W^\pm$ propagators are not shown.}
    \label{fig:2HDM}
\end{figure}

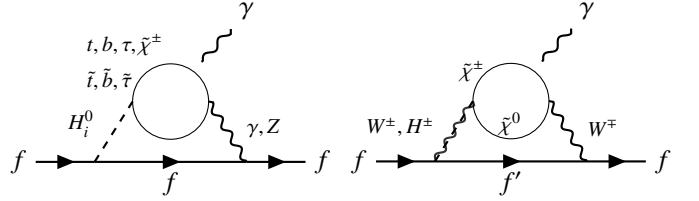
\begin{figure}[t]
    \begin{subfigure}{0.18\textwidth}
        \centering
        \begin{tikzpicture}
        \begin{feynman}
               \vertex at (0,0) (a) {\(f\)};
               \vertex at (1,0) (x);
               \vertex at (3, 0) (y);
               \vertex at (4, 0) (b) {\(f\)};
               \vertex[empty dot,shape=ellipse,minimum height=1.0cm,minimum width=1.0cm] at (2,0.8) (V) {};
               \vertex at (2.3,1.2) (d) {};
               \vertex at (3, 2) (c) {\(\gamma\)};
           \diagram*{
               (a) -- [fermion, thick] (x) -- [fermion, edge label' = \(f\), thick] (y) -- [fermion, thick] (b),
               (x) -- [scalar, thick, edge label ={\footnotesize\(H_i^0\)}, near start] (V.west),
               (y) -- [boson, thick, edge label'= {\footnotesize\(\gamma,Z\)}, near start] (V.east),
               (d) -- [photon, thick] (c),
               };
        \end{feynman}
        \node at (1.4,1.3) {\footnotesize$\begin{aligned}
            &t,b,\tau,\tilde{\chi}^\pm\\
            &\tilde{t},\tilde{b},\tilde{\tau}
        \end{aligned}$};
        \end{tikzpicture}
    \end{subfigure}
    \hspace{1cm}
    \begin{subfigure}{0.18\textwidth}
        \centering
        \begin{tikzpicture}
        \begin{feynman}
               \vertex at (0,0) (a) {\(f\)};
               \vertex at (1,0) (x);
               \vertex at (3, 0) (y);
               \vertex at (4, 0) (b) {\(f\)};
               \vertex[empty dot,shape=ellipse,minimum height=1.0cm,minimum width=1.0cm] at (2,0.8) (V) {};
               \vertex at (2.3,1.2) (d) {};
               \vertex at (3, 2) (c) {\(\gamma\)};
           \diagram*{
               (a) -- [fermion, thick] (x) -- [fermion, edge label' = \(f^\prime\), thick] (y) -- [fermion, thick] (b),
               (x) -- [scalar, thick, edge label ={\footnotesize$W^\pm,H^\pm$}, near start] (V.west),
               (x) -- [boson, thick, opacity = 0.8] (V.west),
               (y) -- [boson, thick, edge label'= {\footnotesize\(W^\mp\)}, near start] (V.east),
               (d) -- [photon, thick] (c),
               };
        \end{feynman}
        \node at (2.0, 0.5) {\footnotesize$\tilde{\chi}^0$};
        \node at (1.5, 1.2) {\footnotesize$\tilde{\chi}^\pm$};
        \end{tikzpicture}
    \end{subfigure}
    \caption{Two-loop diagrams contributing to fermion EDM in the MSSM; left: $\gamma H^0_i$ exchange diagram $(d_f^E)^{\gamma H}$ and $ZH^0_i$ exchange diagram $(d_f^E)^{Z H}$, right: $W^\mp H^\pm$ exchange diagram $(d_f^E)^{W H}$ and $W^\mp W^\pm$ exchange diagram $(d_f^E)^{W W}$. The detached photon line can be inserted on internal propagators or vertices.}
    \label{fig:MSSMTwoLoop}
\end{figure}
Analytic expressions for fermion EDMs in the NMSSM have been obtained by direct generalization of MSSM results (see Refs.~\cite{Ellis:2008zy,Cheung:2011wn}), and have since become standard references in the literature. However, we find this generalization to be incomplete, as it overlooks a critical distinction between the two models. Unlike the MSSM, which lacks tree-level CP violation in the Higgs sector—its only CP-violating phase, $\phi_0$, vanishes trivially—the NMSSM permits such violation at tree level. As a result, one must incorporate the characteristic two-loop diagrams familiar from the 2HDM that are sensitive to CP-violating Higgs-sector phases. Fortunately, the type-II 2HDM closely resembles the NMSSM in its Higgs couplings, making it feasible to adapt 2HDM-based EDM analyses to the NMSSM with only moderate modifications. For instance, gauge-invariant contributions to light-fermion EDMs from Barr--Zee diagrams with the $H^\pm$- and $W^\pm$-upper loops have been thoroughly investigated in the 2HDM ~\cite{Abe:2013qla,Inoue:2014nva}, and more recent work has shown that the ``Kite" diagram also yields a non-negligible contribution~\cite{Altmannshofer:2020shb}.
Additionally, Refs.~\cite{Davila:2025goc,Altmannshofer:2025nsl} propose new sources of electron EDMs arising from $Z_2$-violating couplings in the charged-Higgs sector of the aligned 2HDM. However, such CP-violating couplings are absent in the NMSSM, where the $H^+ \bar{f}^\prime f$ vertices are identical to those in the type-II 2HDM. Fig.~\ref{fig:2HDM} illustrates the type-II-2HDM diagrams  relevant to light-fermion EDMs in the NMSSM—beyond the standard Barr–Zee diagrams—in the background field gauge, and ~\ref{sec:2HDMformulae} outlines the necessary steps to obtain the corresponding analytic expressions. We note that while sfermions can, in principle, enter the internal loops of the Kite diagrams, their contributions are strongly suppressed by heavy sparticle masses and are thus neglected in our analysis. 

It is natural to categorize the electron EDM contributions into three distinct sources: the one-loop MSSM contribution (Fig.~\ref{fig:MSSMOneLoop}), the two-loop MSSM contribution (Fig.~\ref{fig:MSSMTwoLoop}), and the 2HDM contribution (Fig.~\ref{fig:2HDM}). Each group of contributions is gauge invariant itself with gauge dependence canceled out internally. Table~\ref{Tab:DiagramPhaseDependence} summarizes the sensitivity of each subgroup of diagrams to the CP-violating phases, distinguishing those that are specific to the MSSM from those that arise solely within the extended NMSSM structure. This classification highlights whether the contributions arise from the minimal supersymmetric framework or require the extended structure of the NMSSM.

\section{Numerical study}
\label{sec:numerical study}
In this section, we examine how the inclusion of 2HDM diagrams affects the constraints on CP-violating phases from electron EDM measurements. Our procedure is as follows. First, we choose input parameters and solve the tadpole equations for $m^2_{H_d}$, $m^2_{H_u}$, $m^2_S$, $\phi_7$, and $\phi_8$. Using these solutions, we determine the mass spectrum and CP-violating couplings, which are then linked to the electron EDM via analytic expressions. Although our analysis treats the Higgs potential at tree level, we confirm the viability of our benchmark points with \texttt{NMSSMTools}~\cite{Ellwanger:2005dv,Ellwanger:2004xm} in the CP-conserving limit, ensuring the absence of Landau poles, false vacua, and conflicts with LEP, Tevatron, and LHC exclusion data~\cite{ALEPH:2006tnd,Davies:2013aua,ATLAS:2019nkf}. 

To gain insights into the electron EDM's sensitivity to individual CP-violating phases, we first analyze a representative benchmark point, \textbf{BP}. The constraints on \textbf{BP} imposed by the electron EDM will illuminate our understanding of the broader scan performed later on the Higgs sector parameters. The benchmark point \textbf{BP} is defined by
\begin{equation}
\begin{aligned}
    &\tan\beta = 2.11, ~~\lambda = 0.65, ~~\kappa=0.2,\\
    &v_s = 494 \text{GeV}, ~~ \alpha = 6.34\times 10^4 \text{GeV}^2, ~~ \beta = 200\text{GeV}
\end{aligned}
\end{equation}
with the following soft SUSY breaking couplings and masses
\begin{equation}
    \begin{aligned}
        & M_2 = 2M_1 = 200\text{GeV}, ~~ M_3 = 2.5\text{TeV}, \\
        &b = 1.0 \times 10^4 \text{GeV}^2, ~~ A_u = A_e = A_d =200\text{GeV}, \\
        &m_7^2 = 5.56 \times 10^4 \text{GeV}^2, ~~m_9^3 = 9.82 \times 10^6 \text{GeV}^3, \\
        & A_\lambda = 959\text{GeV}, ~ A_\kappa = 900\text{GeV},~~ \mu^\prime =200\text{GeV}.
    \end{aligned}
\end{equation}
We set the masses of the third-generation fermions to $1$TeV, while those of the first two generations are fixed at $5$TeV. The central values of invariant CP-violating phases are chosen as
\begin{equation}
    \Phi_{1,2,3,4,5} = \phi_{0,1,2,4,6,8,9} = 0, ~~\phi_0^\prime = \phi_{3,5,7} = \pi.
\end{equation}
\begin{figure}[t]
    \centering
    \includegraphics[width=1.0\linewidth]{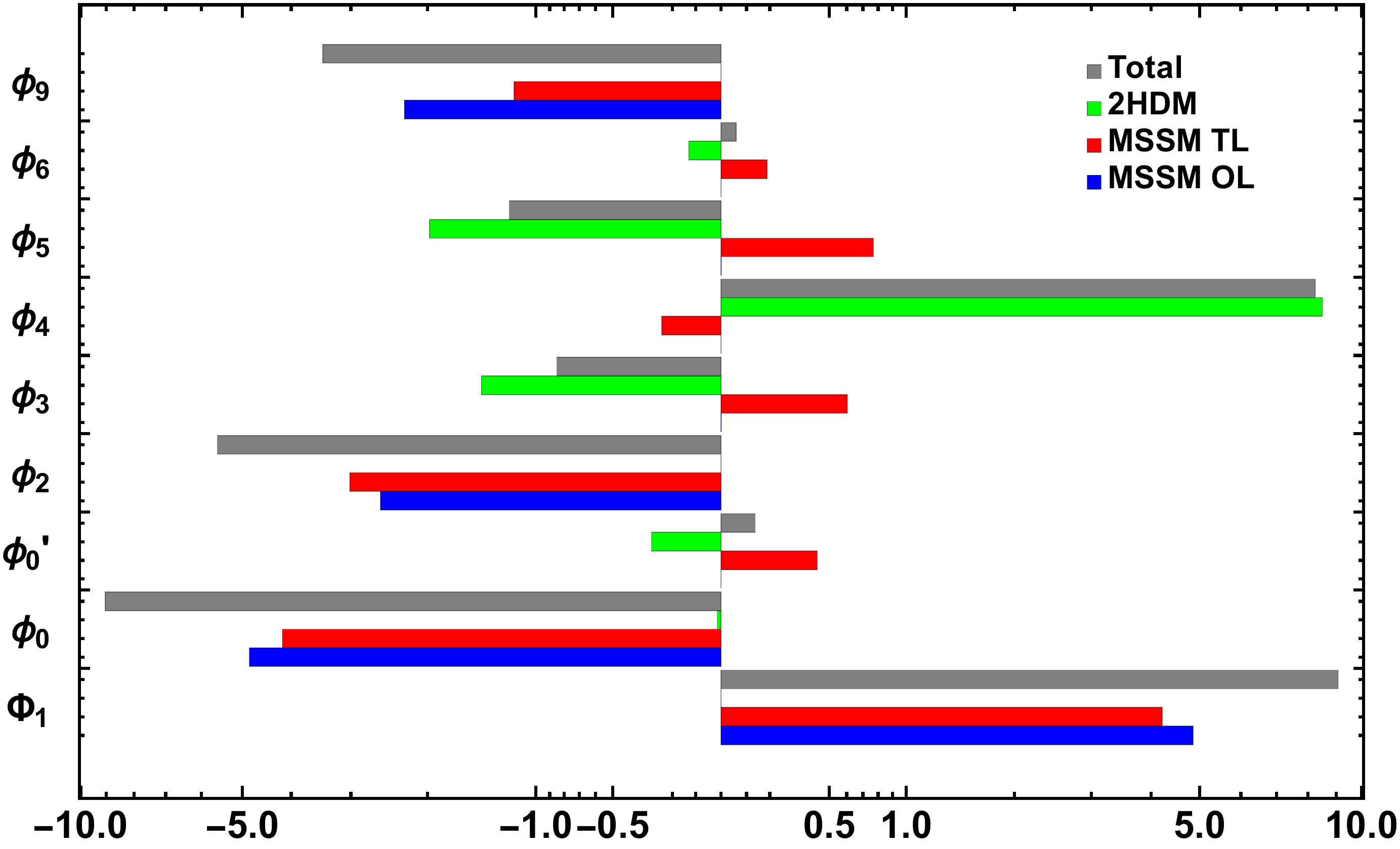}
    \caption{Magnitudes of three groups of electron EDM diagrams caused by individual CP-violating phases with $
    \sin(\phi) = 0.01$. The lengths of bars are shown in unit of  $d_e^{\text{Exp}} = 4.1 \times 10^{-30}$ e cm. Parameters are explained in text.}
    \label{fig:singlephasecompare}
\end{figure}
Setting each phase to a non-zero value of $\sin(\phi)=0.01$ while holding other parameters fixed, we show the resulting grouped electron EDM contributions and their totals in Fig.~\ref{fig:singlephasecompare}. We omit $\Phi_{3,4}$ and $\phi_1$ because their effects on 
$(d_f^E)^{\gamma H}$ and $(d_f^E)^{\tilde{\chi}^0}$ are strongly suppressed by the heavy masses of stops, sbottoms, and selectrons. In contrast, the MSSM-specific phases $\Phi_1$, $\phi_{0}$, $\phi_{9}$, together with the ``effective-MSSM phase'' $\phi_2$, yield comparably significant contributions to both MSSM one-loop (OL) and two-loop (TL) diagrams, all with the same sign. This occurs because, even with heavy sneutrinos ($m_{\tilde{\nu}}\sim 5\,\mathrm{TeV}$), the one-loop chargino contribution$(d_f^E)^{\tilde{\chi}^\pm}$ remains highly sensitive to these phases. Conversely, heavy selectrons effectively suppress $(d_f^E)^{\tilde{\chi}^0}$, resulting in negligible MSSM OL contributions from $\phi_3$ and $\phi_5$. These findings are consistent with the observations reported in Ref.~\cite{King:2015oxa}.

The influence of 2HDM diagrams on the electron EDM is phase-dependent. Notably, the 2HDM contributions from $\phi_{3,4,5}$ dominate over MSSM OL and TL effects, not only reversing the EDM sign but also reaching magnitudes comparable to MSSM-specific phases. This highlights the significance of 2HDM contributions, particularly since $\phi_{3}$ and $\phi_{4}$—arising from the $\mathbb{Z}3$-NMSSM—were previously considered weakly constrained by EDM limits~\cite{Cheung:2011wn,King:2015oxa}. Conversely, $\phi_0^\prime$ and $\phi_6$ induce smaller 2HDM contributions, opposite in sign to MSSM TL effects, suggesting potential cancellations that may relax EDM constraints. The relative suppression of EDMs from $\phi_0^\prime$, $\phi_6$, and $\phi_7$ arises from their CP-violating origins in singlet self-couplings. Unlike most NMSSM-specific phases involving Im$(\lambda)$, which enter EDM diagrams directly via tree-level CP violation in $H_u$-$H_d$ couplings, these phases contribute indirectly through intermediate $H_{u,d}$-$S$ and $S$-$S$ interactions. Consequently, their EDM effects are naturally suppressed by additional power of couplings and singlet masses. Finally, $\phi_0$ has negligible impact on 2HDM diagrams compared to its role in MSSM diagrams for this benchmark point. 

\begin{figure}[t]
    \centering
    \includegraphics[width=1.0\linewidth]{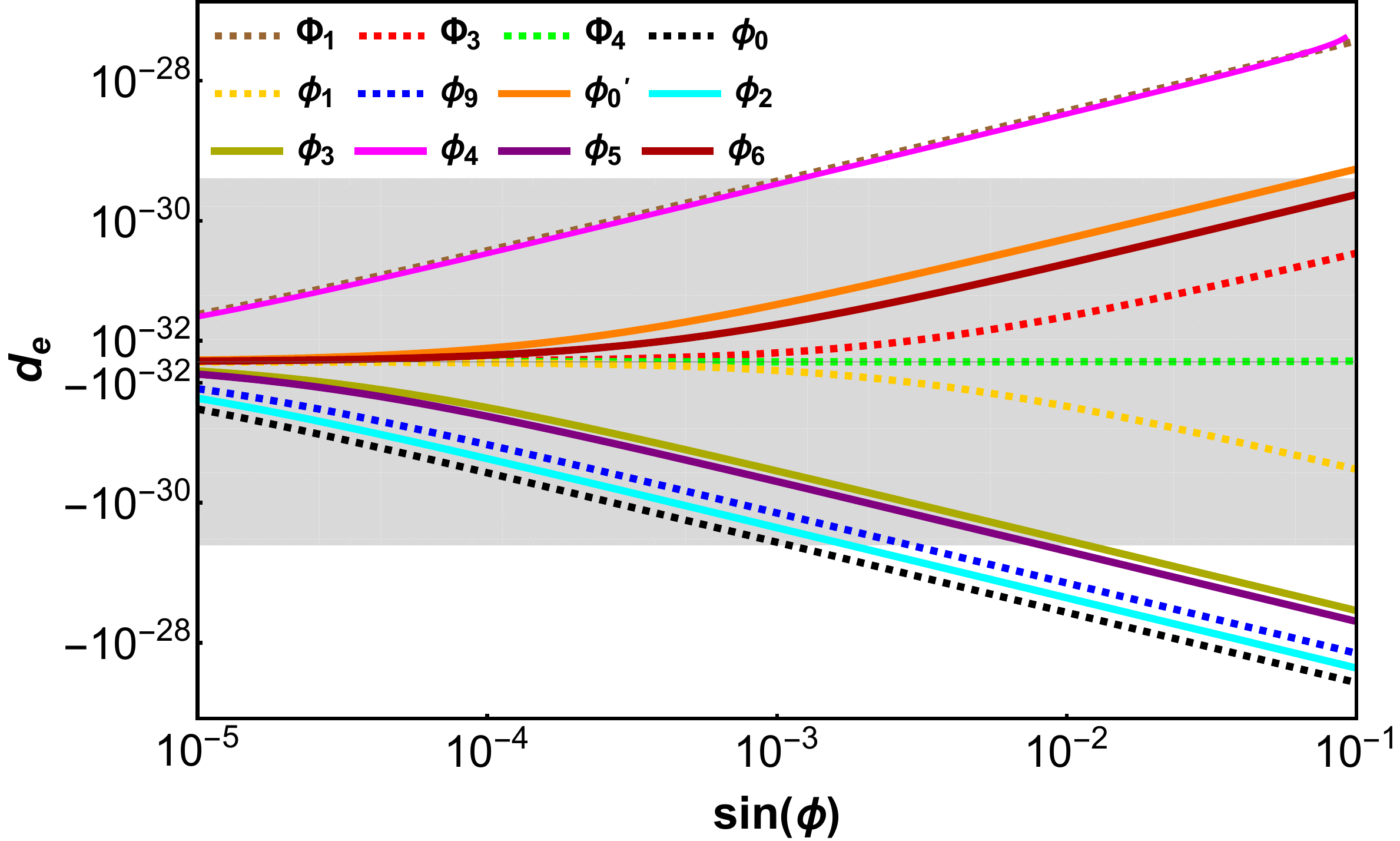}
    \caption{Magnitudes of electron EDM (in unit of e cm) induced by MSSM phases (dash curves) and NMSSM phases (solid curves). The gray region denotes the current experimental bound. Parameters are explained in text.}
    \label{fig:singlephaseconstraint}
\end{figure}

Fig.\ref{fig:singlephaseconstraint} compares the contribution to the electron EDM from each CP-violating phase varied over the range $10^{-5} \sim 10^{-1}$. Phases $\Phi_3$, $\Phi_4$, and $\phi_1$ produce EDMs well below the experimental limit, with $\Phi_4$ being particularly unconstrained due to the small Yukawa 
coupling $y_b$. As noted earlier, cancellations between 2HDM and MSSM diagrams further suppress the already small magnitudes from $\phi_0^\prime$ and $\phi_6$. In contrast, the phases $\phi_3$ and $\phi_5$ are moderately constrained, with $\sin(\phi) \lesssim 10^{-2}$ sufficient to evade the EDM bound. Phases originating in the chargino sector---namely $\Phi_1$ and $\phi_{0,1,2,9}$---face tighter restrictions, demanding $\sin(\phi) \lesssim 10^{-3}$ for this benchmark point. Notably, the inclusion of 2HDM diagrams dramatically alters the constraint on $\phi_4$; whereas $\phi_4$ resembled same EDM magnitude as $\phi_6$ under MSSM-TL contributions alone, it now demands $\sin(\phi_4) \lesssim 10^{-3}$ to remain consistent with EDM measurements. 

Moreover, since the phases $\phi_{3,4,5}$ experience similar modifications from 2HDM contributions, their relative EDM magnitudes---as shown in Figs.~\ref{fig:singlephasecompare} and \ref{fig:singlephaseconstraint}---can be understood by comparing the relevant couplings in 
Eq.~\eqref{eq:NMSSMphases}. In particular, the coupling combination associated with $\phi_3$ is suppressed relative to $\phi_4$ by a factor $\lvert \kappa v_s / A_\lambda \rvert \sim \mathcal{O}(0.1)$, whereas the ratio of coupling combinations for $\phi_5$ to $\phi_3$ is $\lvert \beta / \kappa v_s \rvert \sim \mathcal{O}(1)$. Consequently, $\phi_4$ is constrained roughly an order of magnitude more than $\phi_3$, while $\phi_5$ remains on the same footing as $\phi_3$.

\begin{figure}[t]
    \centering
    \begin{subfigure}{1.0\linewidth}
    \centering
        \includegraphics[width=0.9\linewidth]{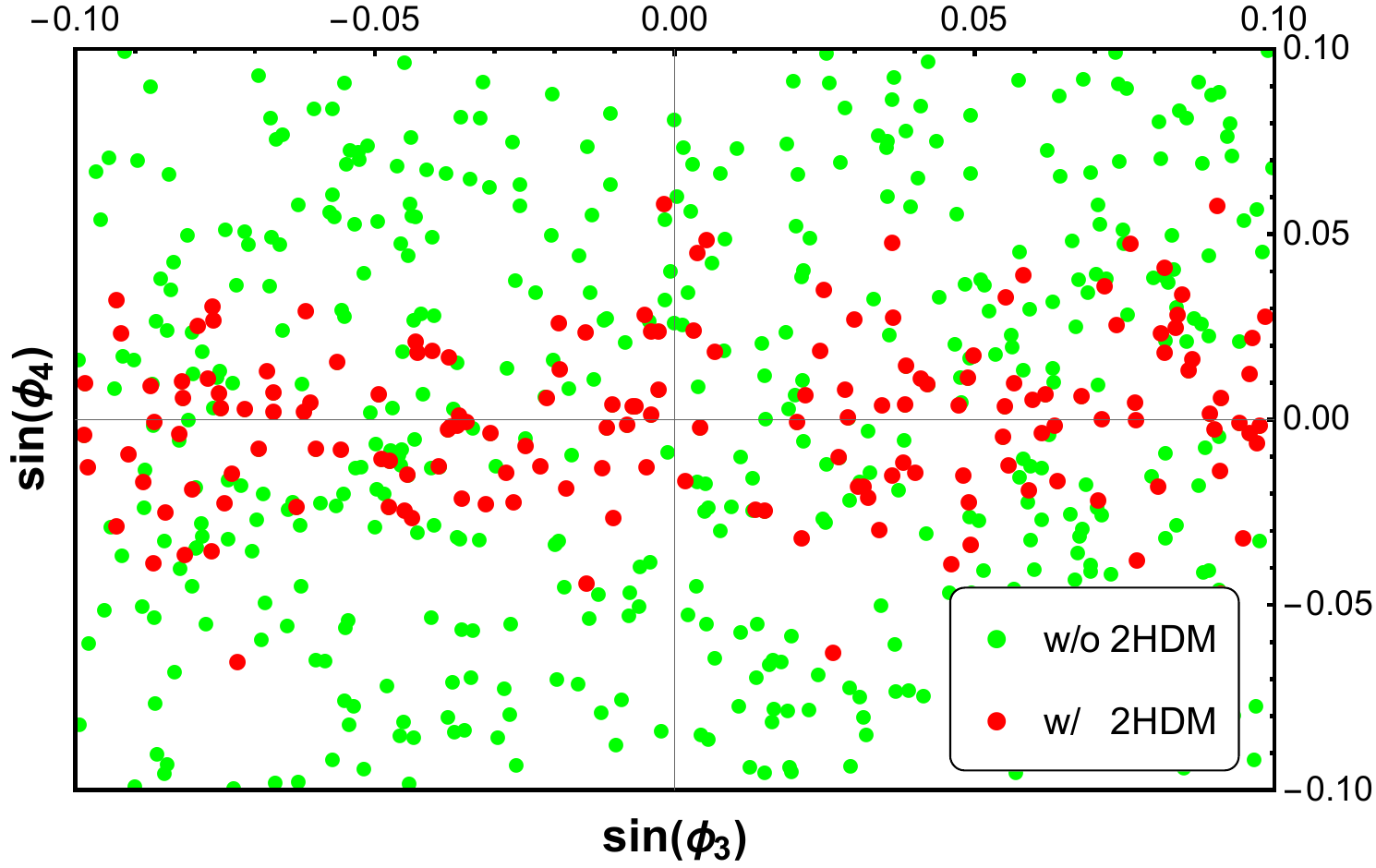}
        \caption{}
        \label{fig:scanphi3phi4}
    \end{subfigure}
    \begin{subfigure}{1.0\linewidth}
    \centering
        \includegraphics[width=0.9\linewidth]{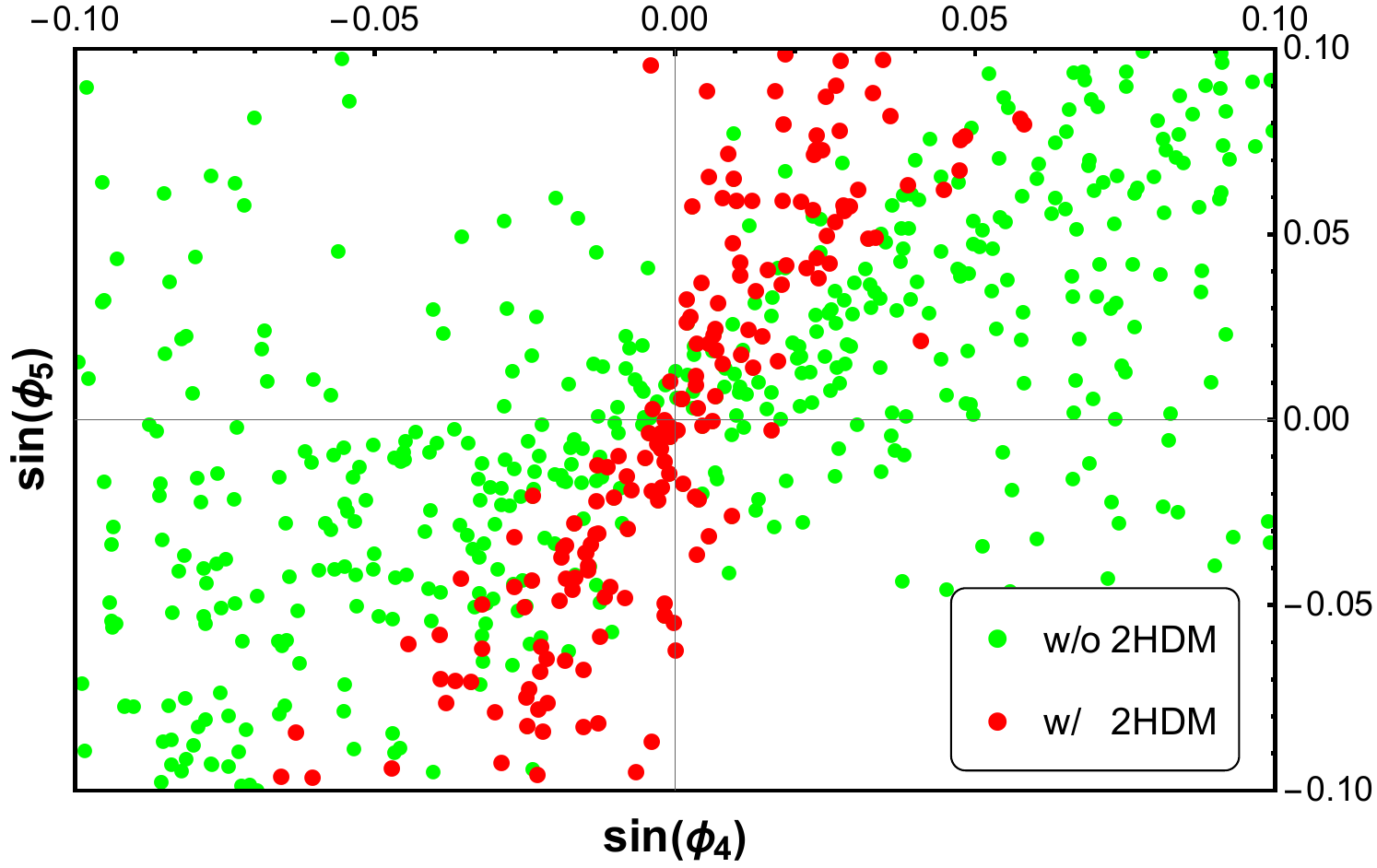}
        \caption{}
        \label{fig:scanphi4phi5}
    \end{subfigure}
    \caption{Parameters points that pass the limit of electron EDM. Red and green dots represent constraints with and without inclusion of 2HDM contributions.}
    \label{fig:scan}
\end{figure}

Next we perform a parameter scan to evaluate the impact of 2HDM loop diagrams on the electron EDM constraints, focusing on the NMSSM CP-violating $\phi_3$, $\phi_4$ and $\phi_5$. We sample the NMSSM parameters within the following ranges:
\begin{equation}
    \begin{aligned}
        &0.5\leq\lambda\leq 0.7, ~ 0.1\leq\kappa\leq 0.3, ~ 200\text{GeV}\leq v_s\leq 500\text{GeV},\\
        & 200\text{GeV}\leq \sqrt{\alpha}\leq 500\text{GeV},~ 200\text{GeV}\leq \beta\leq 1\text{TeV},\\
        &350\text{GeV}\leq A_\lambda, A_\kappa\leq 1\text{TeV}, ~100\text{GeV}\leq m_7, m_9\leq 500\text{GeV},
    \end{aligned}
\end{equation}
and retain the MSSM parameter choices outlined in benchmark \textbf{BP}, except for $\tan\beta = 2.0, ~M_2 = 2M_1 = 400\text{GeV}$. 
The phases in the range 
$-0.1 \leq \sin(\phi_{3,4,5}) \leq 0.1$ are scanned. In Fig.\ref{fig:scan}, we compare the allowed parameter space for these phases with and without the inclusion of 2HDM contributions, illustrating their critical role in shaping EDM constraints. 

As illustrated in Fig.~\ref{fig:scan}, 2HDM contributions significantly enhance the sensitivity of the electron EDM to $\phi_4$ compared to $\phi_5$ and $\phi_3$. While $\phi_{3}$ and $\phi_{5}$ can span the entire scanned range without violating EDM constraints, most points with large $\phi_4$ values are ruled out once 2HDM diagrams are included, imposing a stringent limit $|\sin(\phi_4)| \lesssim 0.05$. Figure~\ref{fig:scanphi3phi4} further shows minimal correlation between $\phi_3$ and $\phi_4$ under EDM constraints, mainly because $\phi_3$ contributes too little to significantly affect 
$\phi_4$. By contrast, $\phi_5$ can produce sizable and opposite-sign EDM contributions that partially cancel those from $\phi_4$. Consequently, both with and without 2HDM diagrams, the allowed regions for $\phi_4$ and $\phi_5$ lie predominantly in the first and third quadrants to facilitate 
cancellation. Moreover, with 2HDM contributions included, $\phi_4$ is noticeably more constrained than $\phi_5$, causing the valid points (shown in red) to cluster closer to the vertical axis. In other words, satisfying the current electron EDM limit demands a more fine-tuned interplay between 
$\phi_4$ and $\phi_5$.

\section{Summary and conclusions}
We have identified a class of Feynman diagrams contributing significantly to the electron EDM in the NMSSM that were previously overlooked in the literature. These diagrams, analogous to those considered in 2HDM, incorporate the effects of all CP-violating interactions within the scalar Higgs sector. Starting with a general NMSSM framework, we established a basis of 14 independent CP-violating phases. Among the NMSSM-specific phases defined in Eq.~\eqref{eq:NMSSMphases}, $\phi_3$, $\phi_4$, and $\phi_5$ exhibit the largest 2HDM-type contributions. Moreover, the latter surpass their corresponding MSSM contributions, reversing the predicted signs of their electron EDM effects.  In contrast, the phases $\phi_0^\prime$ and $\phi_6$ produce slightly smaller 2HDM contributions opposite in sign to their MSSM counterparts, while $\phi_0$ induces negligible 2HDM effects. Notably, the $\phi_4$ -- a phase intrinsic to the $\mathbb{Z}_3$-invariant NMSSM (alongside $\phi_0^\prime$ and $\phi_3$) -- becomes significantly more constrained when 2HDM diagrams are included, with sensitivity an order of magnitude stronger than $\phi_3$ and $\phi_5$. Conversely, $\phi_0^\prime$ and $\phi_6$ remain weakly constrained due to cancellations between 2HDM and MSSM contributions.

From a broader perspective, the study of CP violation extends beyond low-energy EDMs to include baryogenesis and high-energy collider measurements, providing a multi-faceted approach to constrain new physics (see, \textit{e.g.}, Refs~\cite{Chen:2017com,Cirigliano:2009yd}). In this context, our work, providing revised formulae for elementary fermion EDMs and updated constraints on individual NMSSM CP-violating phases, offers a critical starting point for subsequent research in baryogenesis and collider phenomenology within the NMSSM~\cite{Kaifei2025}. Additionally, investigations into EDMs of other composite systems (\textit{e.g.}, atoms and molecules) are indispensable, as their unique dependence on CP-violating phases yields distinct sources of experimental bounds. Ultimately, a comprehensive approach integrating all such sensitive phenomena is crucial for pushing the boundaries of CP-violation tests in BSM scenarios.
\section*{Acknowledgment}
We thank J. Balal Habashi for useful discussions during the initial stages of this work. This research was supported in part under U.S. Department of Energy contract no. DE-SC0011095.

\appendix
\section{Extending 2HDM Results to the NMSSM}
\label{sec:2HDMformulae}
Ref.~\cite{Altmannshofer:2020shb} provides a comprehensive two-loop calculation of the electron EDM and achieves a fully gauge-invariant result in the 2HDM. In what follows, we adapt those findings into the conventional SUSY notation. First, we define the neutral Higgs couplings to quarks and charged leptons as
\begin{equation}
\label{eq:YukawaCoupling}
    \mathcal{L}_{H_i\bar{f}f} = -\frac{m_f}{v}\sum_i H_i \bar{f}\left(g^S_{H_i\bar{f}f} + i g^P_{H_i\bar{f}f} \gamma_5\right)f
\end{equation}
where $v =\sqrt{v_u^2 + v_d^2}$. The following 
couplings of neutral Higgs bosons to charged Higgs bosons and gauge bosons 
are also relevant:
\begin{equation}
\label{eq:HiggsGaugeChargedH}
    \mathcal{L} \supset \sum_i 2\frac{m_W^2}{v} a_i W_\mu^+ W^{-\mu} H_i + \frac{m_Z^2}{v} a_i Z_\mu Z^{\mu} H_i - v\lambda_i H_i^0 H^+ H^-
\end{equation}
Let the neutral scalar fields be rotated into the mass-eigenstate basis via $(h_d, h_u, h_s, a, a_s)_i ={O}_{ij}H_j^0$ where $a$ stands for the CP-odd Higgs field. At tree level, the couplings appearing in Eqs.~\eqref{eq:YukawaCoupling} and~\eqref{eq:HiggsGaugeChargedH} can be expressed as,
\begin{equation}
    \begin{aligned}
        &(g^S, g^P) = \left\{\begin{aligned}
            &({O}_{1i}/\cos\beta, -{O}_{4i}\tan\beta), ~~ f = l,d\\
            &({O}_{2i}/\sin\beta, -{O}_{4i}\cot\beta), ~~ f = u
        \end{aligned}\right. \\
        &a_i = O_{1i}\cos\beta + O_{2i}\sin\beta \\
        &\lambda_i = O_{1i}g_{h_d H^+ H^-} + O_{2i}g_{h_u H^+ H^-} \\
        &~~~~~~~~~~~~+ O_{3i}g_{h_s H^+ H^-} + O_{5i}g_{a_s H^+ H^-}
    \end{aligned}
\end{equation}
with 
\begin{equation}
    \begin{aligned}
       & g_{h_d H^+ H^-}= \frac{1}{4} \left(g^2-{g^\prime}^2\right) \cos\beta^3 +\frac{1}{2} g^2\cos\beta \sin\beta^2
       \\
       &~~~~~~~~~~~~~~+\frac{1}{4} \left(g^2+{g^\prime}^2\right) \cos\beta \sin\beta^2-|\lambda|^2\cos\beta
        \sin\beta^2 \\
       &g_{h_u H^+ H^-}=  \frac{1}{4} \left(g^2-{g^\prime}^2\right)\sin\beta^3 +\frac{1}{2} 
       g^2 \sin\beta\cos\beta^2\\
       &~~~~~~~~~~~~~~ +\frac{1}{4}\left(g^2+{g^\prime}^2\right) \sin\beta\cos\beta^2 - |\lambda|^2 \sin\beta\cos\beta^2\\
       &vg_{h_s H^+ H^-} = \sqrt{2} |\lambda A_{\lambda}| \cos\beta  \sin\beta \cos \phi_4 \\
       &~~~~~~~~~~~~~~~~+\sqrt{2} |\lambda\beta| 
       \cos\beta \sin\beta \cos\phi_5 \\
       &~~~~~~~~~~~~~~~~+ 2  v_s | \lambda\kappa|
       \sin\beta \cos\beta\cos\phi_3+|\lambda|^2 v_s\\
       &vg_{a_s H^+ H^-} = -\sqrt{2} |\lambda A_{\lambda}| \cos\beta\sin\beta \sin\phi_4 \\
       &~~~~~~~~~~~~~~~~~+ \sqrt{2} |\lambda \beta|
       \cos\beta \sin\beta \sin\phi_5\\
       &~~~~~~~~~~~~~~~~~+2 v_s |\lambda\kappa |
       \sin\beta\cos\beta  \sin \phi_3
    \end{aligned}
\end{equation}
Then, by making the following substitutions in the results of Ref.~\cite{Altmannshofer:2020shb} for the charged Higgs loop \((d_f^E)^{H^\pm}\), the \(W\)-boson loop \((d_f^E)^{W^\pm}\), and the Kite diagram \((d_f^E)^{\text{Kite}}\), one can directly adapt the 2HDM expressions to the NMSSM framework:
\begin{equation}
    q_{i1} \rightarrow a_i, \qquad c_f \text{Im}(q_{i2}) \rightarrow g^P_{H_i \bar{f}\bar{f}}, \qquad \lambda_{iH^+H^-} \rightarrow \lambda_i
\end{equation}

\bibliographystyle{elsarticle-num} 
\bibliography{edm}

\begin{thebibliography}{10}
\expandafter\ifx\csname url\endcsname\relax
  \def\url#1{\texttt{#1}}\fi
\expandafter\ifx\csname urlprefix\endcsname\relax\def\urlprefix{URL }\fi
\expandafter\ifx\csname href\endcsname\relax
  \def\href#1#2{#2} \def\path#1{#1}\fi

\bibitem{ParticleDataGroup:2024cfk}
S.~Navas, et~al., {Review of particle physics}, Phys. Rev. D 110~(3) (2024)
  030001.
\newblock \href {https://doi.org/10.1103/PhysRevD.110.030001}
  {\path{doi:10.1103/PhysRevD.110.030001}}.

\bibitem{Cohen:1993nk}
A.~G. Cohen, D.~B. Kaplan, A.~E. Nelson, {Progress in electroweak
  baryogenesis}, Ann. Rev. Nucl. Part. Sci. 43 (1993) 27--70.
\newblock \href {http://arxiv.org/abs/hep-ph/9302210}
  {\path{arXiv:hep-ph/9302210}}, \href
  {https://doi.org/10.1146/annurev.ns.43.120193.000331}
  {\path{doi:10.1146/annurev.ns.43.120193.000331}}.

\bibitem{Kajantie:1995kf}
K.~Kajantie, M.~Laine, K.~Rummukainen, M.~E. Shaposhnikov, {The Electroweak
  phase transition: A Nonperturbative analysis}, Nucl. Phys. B 466 (1996)
  189--258.
\newblock \href {http://arxiv.org/abs/hep-lat/9510020}
  {\path{arXiv:hep-lat/9510020}}, \href
  {https://doi.org/10.1016/0550-3213(96)00052-1}
  {\path{doi:10.1016/0550-3213(96)00052-1}}.

\bibitem{Delepine:1996vn}
D.~Delepine, J.~M. Gerard, R.~Gonzalez~Felipe, J.~Weyers, {A Light stop and
  electroweak baryogenesis}, Phys. Lett. B 386 (1996) 183--188.
\newblock \href {http://arxiv.org/abs/hep-ph/9604440}
  {\path{arXiv:hep-ph/9604440}}, \href
  {https://doi.org/10.1016/0370-2693(96)00921-5}
  {\path{doi:10.1016/0370-2693(96)00921-5}}.

\bibitem{Huet:1995sh}
P.~Huet, A.~E. Nelson, {Electroweak baryogenesis in supersymmetric models},
  Phys. Rev. D 53 (1996) 4578--4597.
\newblock \href {http://arxiv.org/abs/hep-ph/9506477}
  {\path{arXiv:hep-ph/9506477}}, \href
  {https://doi.org/10.1103/PhysRevD.53.4578}
  {\path{doi:10.1103/PhysRevD.53.4578}}.

\bibitem{Cline:1997vk}
J.~M. Cline, M.~Joyce, K.~Kainulainen, {Supersymmetric electroweak baryogenesis
  in the WKB approximation}, Phys. Lett. B 417 (1998) 79--86, [Erratum:
  Phys.Lett.B 448, 321--321 (1999)].
\newblock \href {http://arxiv.org/abs/hep-ph/9708393}
  {\path{arXiv:hep-ph/9708393}}, \href
  {https://doi.org/10.1016/S0370-2693(97)01361-0}
  {\path{doi:10.1016/S0370-2693(97)01361-0}}.

\bibitem{Carena:1996wj}
M.~Carena, M.~Quiros, C.~E.~M. Wagner, {Opening the window for electroweak
  baryogenesis}, Phys. Lett. B 380 (1996) 81--91.
\newblock \href {http://arxiv.org/abs/hep-ph/9603420}
  {\path{arXiv:hep-ph/9603420}}, \href
  {https://doi.org/10.1016/0370-2693(96)00475-3}
  {\path{doi:10.1016/0370-2693(96)00475-3}}.

\bibitem{Cline:2000nw}
J.~M. Cline, M.~Joyce, K.~Kainulainen, {Supersymmetric electroweak
  baryogenesis}, JHEP 07 (2000) 018.
\newblock \href {http://arxiv.org/abs/hep-ph/0006119}
  {\path{arXiv:hep-ph/0006119}}, \href
  {https://doi.org/10.1088/1126-6708/2000/07/018}
  {\path{doi:10.1088/1126-6708/2000/07/018}}.

\bibitem{Lee:2004we}
C.~Lee, V.~Cirigliano, M.~J. Ramsey-Musolf, {Resonant relaxation in electroweak
  baryogenesis}, Phys. Rev. D 71 (2005) 075010.
\newblock \href {http://arxiv.org/abs/hep-ph/0412354}
  {\path{arXiv:hep-ph/0412354}}, \href
  {https://doi.org/10.1103/PhysRevD.71.075010}
  {\path{doi:10.1103/PhysRevD.71.075010}}.

\bibitem{Carena:2008rt}
M.~Carena, G.~Nardini, M.~Quiros, C.~E.~M. Wagner, {The Effective Theory of the
  Light Stop Scenario}, JHEP 10 (2008) 062.
\newblock \href {http://arxiv.org/abs/0806.4297} {\path{arXiv:0806.4297}},
  \href {https://doi.org/10.1088/1126-6708/2008/10/062}
  {\path{doi:10.1088/1126-6708/2008/10/062}}.

\bibitem{Kim:1983dt}
J.~E. Kim, H.~P. Nilles, {The mu Problem and the Strong CP Problem}, Phys.
  Lett. B 138 (1984) 150--154.
\newblock \href {https://doi.org/10.1016/0370-2693(84)91890-2}
  {\path{doi:10.1016/0370-2693(84)91890-2}}.

\bibitem{CMS:2021xyz}
C.~Collaboration, Combined searches for the production of supersymmetric top
  quark partners in proton–proton collisions at $\sqrt{s}=13$ tev, CMS
  Physics Analysis SummaryExcludes top squark masses up to 1325 GeV for a
  massless neutralino; 137 fb$^{-1}$ at 13 TeV (2021).
\newblock \href {http://arxiv.org/abs/2107.10892} {\path{arXiv:2107.10892}}.

\bibitem{Chupp:2017rkp}
T.~Chupp, P.~Fierlinger, M.~Ramsey-Musolf, J.~Singh, {Electric dipole moments
  of atoms, molecules, nuclei, and particles}, Rev. Mod. Phys. 91~(1) (2019)
  015001.
\newblock \href {http://arxiv.org/abs/1710.02504} {\path{arXiv:1710.02504}},
  \href {https://doi.org/10.1103/RevModPhys.91.015001}
  {\path{doi:10.1103/RevModPhys.91.015001}}.

\bibitem{ACME:2018yjb}
V.~Andreev, et~al., {Improved limit on the electric dipole moment of the
  electron}, Nature 562~(7727) (2018) 355--360.
\newblock \href {https://doi.org/10.1038/s41586-018-0599-8}
  {\path{doi:10.1038/s41586-018-0599-8}}.

\bibitem{Panico:2018hal}
G.~Panico, A.~Pomarol, M.~Riembau, {EFT approach to the electron Electric
  Dipole Moment at the two-loop level}, JHEP 04 (2019) 090.
\newblock \href {http://arxiv.org/abs/1810.09413} {\path{arXiv:1810.09413}},
  \href {https://doi.org/10.1007/JHEP04(2019)090}
  {\path{doi:10.1007/JHEP04(2019)090}}.

\bibitem{Roussy:2022cmp}
T.~S. Roussy, et~al., {An improved bound on the electron\textquoteright{}s
  electric dipole moment}, Science 381~(6653) (2023) adg4084.
\newblock \href {http://arxiv.org/abs/2212.11841} {\path{arXiv:2212.11841}},
  \href {https://doi.org/10.1126/science.adg4084}
  {\path{doi:10.1126/science.adg4084}}.

\bibitem{Bodeker:2020ghk}
D.~Bodeker, W.~Buchmuller, {Baryogenesis from the weak scale to the grand
  unification scale}, Rev. Mod. Phys. 93~(3) (2021) 035004.
\newblock \href {http://arxiv.org/abs/2009.07294} {\path{arXiv:2009.07294}},
  \href {https://doi.org/10.1103/RevModPhys.93.035004}
  {\path{doi:10.1103/RevModPhys.93.035004}}.

\bibitem{Ross:2016pml}
G.~G. Ross, K.~Schmidt-Hoberg, F.~Staub, {On the MSSM Higgsino mass and fine
  tuning}, Phys. Lett. B 759 (2016) 110--114.
\newblock \href {http://arxiv.org/abs/1603.09347} {\path{arXiv:1603.09347}},
  \href {https://doi.org/10.1016/j.physletb.2016.05.053}
  {\path{doi:10.1016/j.physletb.2016.05.053}}.

\bibitem{Li:2010ax}
Y.~Li, S.~Profumo, M.~Ramsey-Musolf, {A Comprehensive Analysis of Electric
  Dipole Moment Constraints on CP-violating Phases in the MSSM}, JHEP 08 (2010)
  062.
\newblock \href {http://arxiv.org/abs/1006.1440} {\path{arXiv:1006.1440}},
  \href {https://doi.org/10.1007/JHEP08(2010)062}
  {\path{doi:10.1007/JHEP08(2010)062}}.

\bibitem{Han:2021ify}
C.~Han, {Muon g-2 and CP violation in MSSM} (4 2021).
\newblock \href {http://arxiv.org/abs/2104.03292} {\path{arXiv:2104.03292}}.

\bibitem{Nakai:2016atk}
Y.~Nakai, M.~Reece, {Electric Dipole Moments in Natural Supersymmetry}, JHEP 08
  (2017) 031.
\newblock \href {http://arxiv.org/abs/1612.08090} {\path{arXiv:1612.08090}},
  \href {https://doi.org/10.1007/JHEP08(2017)031}
  {\path{doi:10.1007/JHEP08(2017)031}}.

\bibitem{Cesarotti:2018huy}
C.~Cesarotti, Q.~Lu, Y.~Nakai, A.~Parikh, M.~Reece, {Interpreting the Electron
  EDM Constraint}, JHEP 05 (2019) 059.
\newblock \href {http://arxiv.org/abs/1810.07736} {\path{arXiv:1810.07736}},
  \href {https://doi.org/10.1007/JHEP05(2019)059}
  {\path{doi:10.1007/JHEP05(2019)059}}.

\bibitem{Fayet:1974fj}
P.~Fayet, {A Gauge Theory of Weak and Electromagnetic Interactions with
  Spontaneous Parity Breaking}, Nucl. Phys. B 78 (1974) 14--28.
\newblock \href {https://doi.org/10.1016/0550-3213(74)90113-8}
  {\path{doi:10.1016/0550-3213(74)90113-8}}.

\bibitem{Fayet:1974pd}
P.~Fayet, {Supergauge Invariant Extension of the Higgs Mechanism and a Model
  for the electron and Its Neutrino}, Nucl. Phys. B 90 (1975) 104--124.
\newblock \href {https://doi.org/10.1016/0550-3213(75)90636-7}
  {\path{doi:10.1016/0550-3213(75)90636-7}}.

\bibitem{Fayet:1977yc}
P.~Fayet, {Spontaneously Broken Supersymmetric Theories of Weak,
  Electromagnetic and Strong Interactions}, Phys. Lett. B 69 (1977) 489.
\newblock \href {https://doi.org/10.1016/0370-2693(77)90852-8}
  {\path{doi:10.1016/0370-2693(77)90852-8}}.

\bibitem{Ellis:1988er}
J.~R. Ellis, J.~F. Gunion, H.~E. Haber, L.~Roszkowski, F.~Zwirner, {Higgs
  Bosons in a Nonminimal Supersymmetric Model}, Phys. Rev. D 39 (1989) 844.
\newblock \href {https://doi.org/10.1103/PhysRevD.39.844}
  {\path{doi:10.1103/PhysRevD.39.844}}.

\bibitem{Ellwanger:1993xa}
U.~Ellwanger, M.~Rausch~de Traubenberg, C.~A. Savoy, {Particle spectrum in
  supersymmetric models with a gauge singlet}, Phys. Lett. B 315 (1993)
  331--337.
\newblock \href {http://arxiv.org/abs/hep-ph/9307322}
  {\path{arXiv:hep-ph/9307322}}, \href
  {https://doi.org/10.1016/0370-2693(93)91621-S}
  {\path{doi:10.1016/0370-2693(93)91621-S}}.

\bibitem{Ellwanger:2009dp}
U.~Ellwanger, C.~Hugonie, A.~M. Teixeira, {The Next-to-Minimal Supersymmetric
  Standard Model}, Phys. Rept. 496 (2010) 1--77.
\newblock \href {http://arxiv.org/abs/0910.1785} {\path{arXiv:0910.1785}},
  \href {https://doi.org/10.1016/j.physrep.2010.07.001}
  {\path{doi:10.1016/j.physrep.2010.07.001}}.

\bibitem{Maniatis:2009re}
M.~Maniatis, {The Next-to-Minimal Supersymmetric extension of the Standard
  Model reviewed}, Int. J. Mod. Phys. A 25 (2010) 3505--3602.
\newblock \href {http://arxiv.org/abs/0906.0777} {\path{arXiv:0906.0777}},
  \href {https://doi.org/10.1142/S0217751X10049827}
  {\path{doi:10.1142/S0217751X10049827}}.

\bibitem{Cheung:2011wn}
K.~Cheung, T.-J. Hou, J.~S. Lee, E.~Senaha, {Higgs Mediated EDMs in the
  Next-to-MSSM: An Application to Electroweak Baryogenesis}, Phys. Rev. D 84
  (2011) 015002.
\newblock \href {http://arxiv.org/abs/1102.5679} {\path{arXiv:1102.5679}},
  \href {https://doi.org/10.1103/PhysRevD.84.015002}
  {\path{doi:10.1103/PhysRevD.84.015002}}.

\bibitem{Ibrahim:1997gj}
T.~Ibrahim, P.~Nath, {The Neutron and the electron electric dipole moment in
  N=1 supergravity unification}, Phys. Rev. D 57 (1998) 478--488, [Erratum:
  Phys.Rev.D 58, 019901 (1998), Erratum: Phys.Rev.D 60, 079903 (1999), Erratum:
  Phys.Rev.D 60, 119901 (1999)].
\newblock \href {http://arxiv.org/abs/hep-ph/9708456}
  {\path{arXiv:hep-ph/9708456}}, \href
  {https://doi.org/10.1103/PhysRevD.58.019901}
  {\path{doi:10.1103/PhysRevD.58.019901}}.

\bibitem{Barr:1990vd}
S.~M. Barr, A.~Zee, {Electric Dipole Moment of the Electron and of the
  Neutron}, Phys. Rev. Lett. 65 (1990) 21--24, [Erratum: Phys.Rev.Lett. 65,
  2920 (1990)].
\newblock \href {https://doi.org/10.1103/PhysRevLett.65.21}
  {\path{doi:10.1103/PhysRevLett.65.21}}.

\bibitem{Chang:2005ac}
D.~Chang, W.-F. Chang, W.-Y. Keung, {Electric dipole moment in the split
  supersymmetry models}, Phys. Rev. D 71 (2005) 076006.
\newblock \href {http://arxiv.org/abs/hep-ph/0503055}
  {\path{arXiv:hep-ph/0503055}}, \href
  {https://doi.org/10.1103/PhysRevD.71.076006}
  {\path{doi:10.1103/PhysRevD.71.076006}}.

\bibitem{Li:2008kz}
Y.~Li, S.~Profumo, M.~Ramsey-Musolf, {Higgs-Higgsino-Gaugino Induced Two Loop
  Electric Dipole Moments}, Phys. Rev. D 78 (2008) 075009.
\newblock \href {http://arxiv.org/abs/0806.2693} {\path{arXiv:0806.2693}},
  \href {https://doi.org/10.1103/PhysRevD.78.075009}
  {\path{doi:10.1103/PhysRevD.78.075009}}.

\bibitem{Ellis:2008zy}
J.~R. Ellis, J.~S. Lee, A.~Pilaftsis, {Electric Dipole Moments in the MSSM
  Reloaded}, JHEP 10 (2008) 049.
\newblock \href {http://arxiv.org/abs/0808.1819} {\path{arXiv:0808.1819}},
  \href {https://doi.org/10.1088/1126-6708/2008/10/049}
  {\path{doi:10.1088/1126-6708/2008/10/049}}.

\bibitem{King:2015oxa}
S.~F. King, M.~Muhlleitner, R.~Nevzorov, K.~Walz, {Exploring the CP-violating
  NMSSM: EDM Constraints and Phenomenology}, Nucl. Phys. B 901 (2015) 526--555.
\newblock \href {http://arxiv.org/abs/1508.03255} {\path{arXiv:1508.03255}},
  \href {https://doi.org/10.1016/j.nuclphysb.2015.11.003}
  {\path{doi:10.1016/j.nuclphysb.2015.11.003}}.

\bibitem{Dao:2022rui}
T.~N. Dao, D.~N. Le, M.~M\"uhlleitner, {Leptonic anomalous magnetic and
  electric dipole moments in the CP-violating NMSSM with and without inverse
  seesaw mechanism}, Eur. Phys. J. C 82~(10) (2022) 954.
\newblock \href {http://arxiv.org/abs/2207.12618} {\path{arXiv:2207.12618}},
  \href {https://doi.org/10.1140/epjc/s10052-022-10928-3}
  {\path{doi:10.1140/epjc/s10052-022-10928-3}}.

\bibitem{Pospelov:2005pr}
M.~Pospelov, A.~Ritz, {Electric dipole moments as probes of new physics},
  Annals Phys. 318 (2005) 119--169.
\newblock \href {http://arxiv.org/abs/hep-ph/0504231}
  {\path{arXiv:hep-ph/0504231}}, \href
  {https://doi.org/10.1016/j.aop.2005.04.002}
  {\path{doi:10.1016/j.aop.2005.04.002}}.

\bibitem{Engel:2013lsa}
J.~Engel, M.~J. Ramsey-Musolf, U.~van Kolck, {Electric Dipole Moments of
  Nucleons, Nuclei, and Atoms: The Standard Model and Beyond}, Prog. Part.
  Nucl. Phys. 71 (2013) 21--74.
\newblock \href {http://arxiv.org/abs/1303.2371} {\path{arXiv:1303.2371}},
  \href {https://doi.org/10.1016/j.ppnp.2013.03.003}
  {\path{doi:10.1016/j.ppnp.2013.03.003}}.

\bibitem{Lopez-Fogliani:2005vcg}
D.~E. Lopez-Fogliani, C.~Munoz, {Proposal for a Supersymmetric Standard Model},
  Phys. Rev. Lett. 97 (2006) 041801.
\newblock \href {http://arxiv.org/abs/hep-ph/0508297}
  {\path{arXiv:hep-ph/0508297}}, \href
  {https://doi.org/10.1103/PhysRevLett.97.041801}
  {\path{doi:10.1103/PhysRevLett.97.041801}}.

\bibitem{Lopez-Fogliani:2020gzo}
D.~E. Lopez-Fogliani, C.~Munoz, {Searching for supersymmetry: the $\mu\nu$SSM:
  A short review}, Eur. Phys. J. ST 229~(21) (2020) 3263--3301.
\newblock \href {http://arxiv.org/abs/2009.01380} {\path{arXiv:2009.01380}},
  \href {https://doi.org/10.1140/epjst/e2020-000114-9}
  {\path{doi:10.1140/epjst/e2020-000114-9}}.

\bibitem{Chung:2003fi}
D.~J.~H. Chung, L.~L. Everett, G.~L. Kane, S.~F. King, J.~D. Lykken, L.-T.
  Wang, {The Soft supersymmetry breaking Lagrangian: Theory and applications},
  Phys. Rept. 407 (2005) 1--203.
\newblock \href {http://arxiv.org/abs/hep-ph/0312378}
  {\path{arXiv:hep-ph/0312378}}, \href
  {https://doi.org/10.1016/j.physrep.2004.08.032}
  {\path{doi:10.1016/j.physrep.2004.08.032}}.

\bibitem{Haber:1997if}
H.~E. Haber, {The Status of the minimal supersymmetric standard model and
  beyond}, Nucl. Phys. B Proc. Suppl. 62 (1998) 469--484.
\newblock \href {http://arxiv.org/abs/hep-ph/9709450}
  {\path{arXiv:hep-ph/9709450}}, \href
  {https://doi.org/10.1016/S0920-5632(97)00688-9}
  {\path{doi:10.1016/S0920-5632(97)00688-9}}.

\bibitem{Abe:2013qla}
T.~Abe, J.~Hisano, T.~Kitahara, K.~Tobioka, {Gauge invariant Barr-Zee type
  contributions to fermionic EDMs in the two-Higgs doublet models}, JHEP 01
  (2014) 106, [Erratum: JHEP 04, 161 (2016)].
\newblock \href {http://arxiv.org/abs/1311.4704} {\path{arXiv:1311.4704}},
  \href {https://doi.org/10.1007/JHEP01(2014)106}
  {\path{doi:10.1007/JHEP01(2014)106}}.

\bibitem{Inoue:2014nva}
S.~Inoue, M.~J. Ramsey-Musolf, Y.~Zhang, {CP-violating phenomenology of flavor
  conserving two Higgs doublet models}, Phys. Rev. D 89~(11) (2014) 115023.
\newblock \href {http://arxiv.org/abs/1403.4257} {\path{arXiv:1403.4257}},
  \href {https://doi.org/10.1103/PhysRevD.89.115023}
  {\path{doi:10.1103/PhysRevD.89.115023}}.

\bibitem{Altmannshofer:2020shb}
W.~Altmannshofer, S.~Gori, N.~Hamer, H.~H. Patel, {Electron EDM in the complex
  two-Higgs doublet model}, Phys. Rev. D 102~(11) (2020) 115042.
\newblock \href {http://arxiv.org/abs/2009.01258} {\path{arXiv:2009.01258}},
  \href {https://doi.org/10.1103/PhysRevD.102.115042}
  {\path{doi:10.1103/PhysRevD.102.115042}}.

\bibitem{Davila:2025goc}
J.~M. D{\'a}vila, A.~Karan, E.~Passemar, A.~Pich, L.~Vale~Silva, {The Electric
  Dipole Moment of the electron in the decoupling limit of the aligned
  Two-Higgs Doublet Model} (4 2025).
\newblock \href {http://arxiv.org/abs/2504.16700} {\path{arXiv:2504.16700}}.

\bibitem{Altmannshofer:2025nsl}
W.~Altmannshofer, B.~Assi, J.~Brod, N.~Hamer, J.~Julio, P.~Uttayarat,
  D.~Volkov, {Electron EDM and {\ensuremath{\Gamma}}({\ensuremath{\mu}}
  {\textrightarrow} e{\ensuremath{\gamma}}) in the 2HDM}, JHEP 06 (2025) 156.
\newblock \href {http://arxiv.org/abs/2410.17313} {\path{arXiv:2410.17313}},
  \href {https://doi.org/10.1007/JHEP06(2025)156}
  {\path{doi:10.1007/JHEP06(2025)156}}.

\bibitem{Ellwanger:2005dv}
U.~Ellwanger, C.~Hugonie, {NMHDECAY 2.0: An Updated program for sparticle
  masses, Higgs masses, couplings and decay widths in the NMSSM}, Comput. Phys.
  Commun. 175 (2006) 290--303.
\newblock \href {http://arxiv.org/abs/hep-ph/0508022}
  {\path{arXiv:hep-ph/0508022}}, \href
  {https://doi.org/10.1016/j.cpc.2006.04.004}
  {\path{doi:10.1016/j.cpc.2006.04.004}}.

\bibitem{Ellwanger:2004xm}
U.~Ellwanger, J.~F. Gunion, C.~Hugonie, {NMHDECAY: A Fortran code for the Higgs
  masses, couplings and decay widths in the NMSSM}, JHEP 02 (2005) 066.
\newblock \href {http://arxiv.org/abs/hep-ph/0406215}
  {\path{arXiv:hep-ph/0406215}}, \href
  {https://doi.org/10.1088/1126-6708/2005/02/066}
  {\path{doi:10.1088/1126-6708/2005/02/066}}.

\bibitem{ALEPH:2006tnd}
S.~Schael, et~al., {Search for neutral MSSM Higgs bosons at LEP}, Eur. Phys. J.
  C 47 (2006) 547--587.
\newblock \href {http://arxiv.org/abs/hep-ex/0602042}
  {\path{arXiv:hep-ex/0602042}}, \href
  {https://doi.org/10.1140/epjc/s2006-02569-7}
  {\path{doi:10.1140/epjc/s2006-02569-7}}.

\bibitem{Davies:2013aua}
G.~J. Davies, {Higgs boson searches at the Tevatron}, Front. Phys. (Beijing) 8
  (2013) 270--284.
\newblock \href {https://doi.org/10.1007/s11467-013-0293-0}
  {\path{doi:10.1007/s11467-013-0293-0}}.

\bibitem{ATLAS:2019nkf}
G.~Aad, et~al., {Combined measurements of Higgs boson production and decay
  using up to $80$ fb$^{-1}$ of proton-proton collision data at $\sqrt{s}=$ 13
  TeV collected with the ATLAS experiment}, Phys. Rev. D 101~(1) (2020) 012002.
\newblock \href {http://arxiv.org/abs/1909.02845} {\path{arXiv:1909.02845}},
  \href {https://doi.org/10.1103/PhysRevD.101.012002}
  {\path{doi:10.1103/PhysRevD.101.012002}}.

\bibitem{Chen:2017com}
C.-Y. Chen, H.-L. Li, M.~Ramsey-Musolf, {CP-Violation in the Two Higgs Doublet
  Model: from the LHC to EDMs}, Phys. Rev. D 97~(1) (2018) 015020.
\newblock \href {http://arxiv.org/abs/1708.00435} {\path{arXiv:1708.00435}},
  \href {https://doi.org/10.1103/PhysRevD.97.015020}
  {\path{doi:10.1103/PhysRevD.97.015020}}.

\bibitem{Cirigliano:2009yd}
V.~Cirigliano, Y.~Li, S.~Profumo, M.~J. Ramsey-Musolf, {MSSM Baryogenesis and
  Electric Dipole Moments: An Update on the Phenomenology}, JHEP 01 (2010) 002.
\newblock \href {http://arxiv.org/abs/0910.4589} {\path{arXiv:0910.4589}},
  \href {https://doi.org/10.1007/JHEP01(2010)002}
  {\path{doi:10.1007/JHEP01(2010)002}}.

\bibitem{Kaifei2025}
K.~Ning, M.~Ramsey-Musolf, {Implications of the electron EDM on NMSSM
  electroweak baryogenesis}, in preparation (2025).

\end{thebibliography}






\end{document}